\documentclass[runningheads,a4paper]{llncs}

\title{Simulation leagues: Analysis of competition formats}
\author{David Budden$^{1}$, Peter Wang$^2$, Oliver Obst$^2$, Mikhail Prokopenko$^2$}

\institute{$^1$ The University of Melbourne, Parkville, VIC 3010, Australia \\
$^2$ Statistical Learning, CSIRO Computational Informatics,
        Epping, NSW 1710, Australia\\
        \email{david.budden@unimelb.edu.au}\\
        \email{\{peter.wang,oliver.obst,mikhail.prokopenko\}@csiro.au}}

\date{}

\usepackage[table]{xcolor}
\usepackage{times}
\usepackage{url}

\usepackage{graphicx}
\usepackage{amsmath}

\begin{document}
\maketitle

\begin{abstract}
The selection of an appropriate competition format is critical for both the success and credibility of any competition, both real and simulated. In this paper, the automated parallelism offered by the RoboCupSoccer 2D simulation league is leveraged to conduct a 28,000 game round-robin between the top 8 teams from RoboCup 2012 and 2013. A proposed new competition format is found to reduce variation from the resultant statistically significant team performance rankings by 75\% and 67\%, when compared to the actual competition results from RoboCup 2012 and 2013 respectively. These results are statistically validated by generating 10,000 random tournaments for each of the three considered formats and comparing the respective distributions of ranking discrepancy.
\end{abstract}

\section{Introduction}

\subsection{The RoboCup humanoid challenge}
\label{ss:robocup}

RoboCup (the ``World Cup" of robot soccer) was first proposed in 1997 as a standard problem for the evaluation of theories, algorithms and architectures in areas including artificial intelligence (AI), robotics and computer vision~\cite{kitano1997robocup1}. This proposal followed the observation that traditional AI problems were increasingly unable to meet these requirements and that a new challenge was necessary to initiate the development of next-generation technologies.

The overarching RoboCup goal of developing a team of humanoid robots capable of defeating the FIFA World Cup champion team, coined the ``Millennium Challenge", has proven a major factor in driving research in AI and related areas for over a decade, with a search for the term ``RoboCup" in a major literature database yielding over 23,000 results. Since 1997, researchers and competitors have decomposed this ambitious pursuit into two complementary categories~\cite{kitano1998robocup1}:
\begin{itemize}
\item\textbf{Physical robot league:} Using physical robots to play soccer games. This category now contains many different leagues for both wheeled robots (small-sized~\cite{ssl2013} and mid-sized leagues~\cite{msl2013}) and humanoids (standard platform league~\cite{spl2013} and humanoid league~\cite{humanoid2013}), with each focusing on different aspects of physical robot design~\cite{ha2011development}, motor control and bipedal locomotion~\cite{fountain2013motivated,budden2013probabilistic}, real-time localisation~\cite{annable2013nubugger,budden2013particle} and computer vision~\cite{budden2012novel,budden2013colour,budden2013salient}.
\item\textbf{Software agent league:} Using software or synthetic agents to play soccer games on an official soccer server over a network. This category contains both 2D~\cite{chen7robocup,gliders2013tdp,gliders2014tdp} and 3D~\cite{sim3d2013} simulation leagues.
\end{itemize}
\noindent{The annual RoboCup competition, which attracted 2,500 participants and 40,000 spectators from 40 countries in 2013~\cite{butler2013robocup}, now exhibits a number of non-soccer competitions. The oldest and largest of these, RoboCupRescue, is also separated into physical and simulation leagues~\cite{kitano1998robocup2,kitano2001robocup}.}

\subsection{Significance of simulation leagues}

The RoboCupSoccer simulation leagues traditionally involve the largest number of international participating teams, reaching 40 in 2013~\cite{bai2012wrighteagle}. The ability to simulate soccer matches without physical robots abstracts away low-level issues such as image processing and motor breakages, allowing teams to focus on the development of complex team behaviours and strategies for a larger number of autonomous agents. 
The remainder of this section expands upon some specific contributions of the RoboCupSoccer simulation leagues toward the `Millennium Challenge".

\subsubsection{Financial inclusiveness of competing nations}

\begin{figure}[t]
\begin{center}
\begin{tabular}{>{\centering}p{5.8cm}<{\centering} >{\centering}p{5.8cm}<{\centering}}
\includegraphics[width=5.8cm]{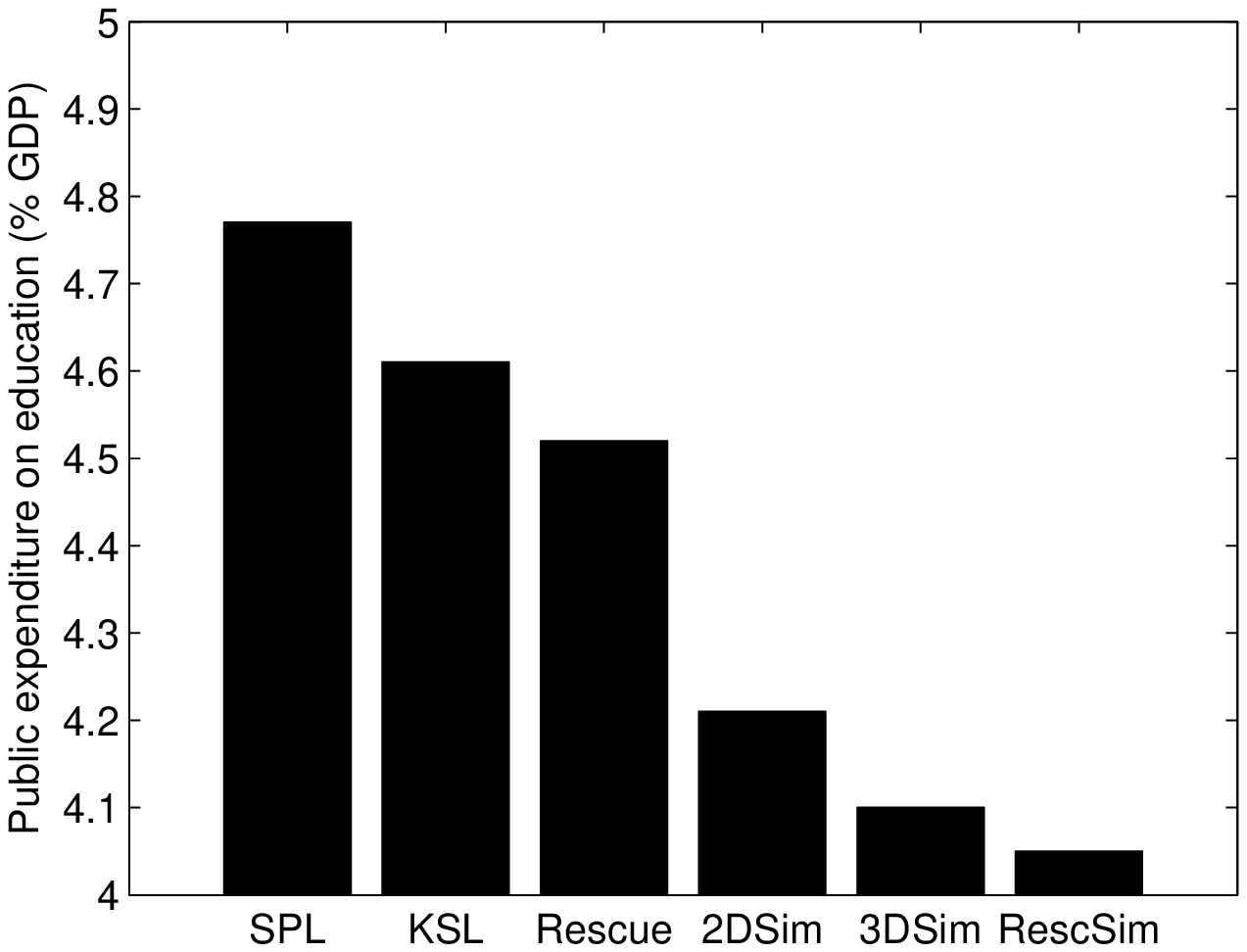} & \includegraphics[width=5.8cm]{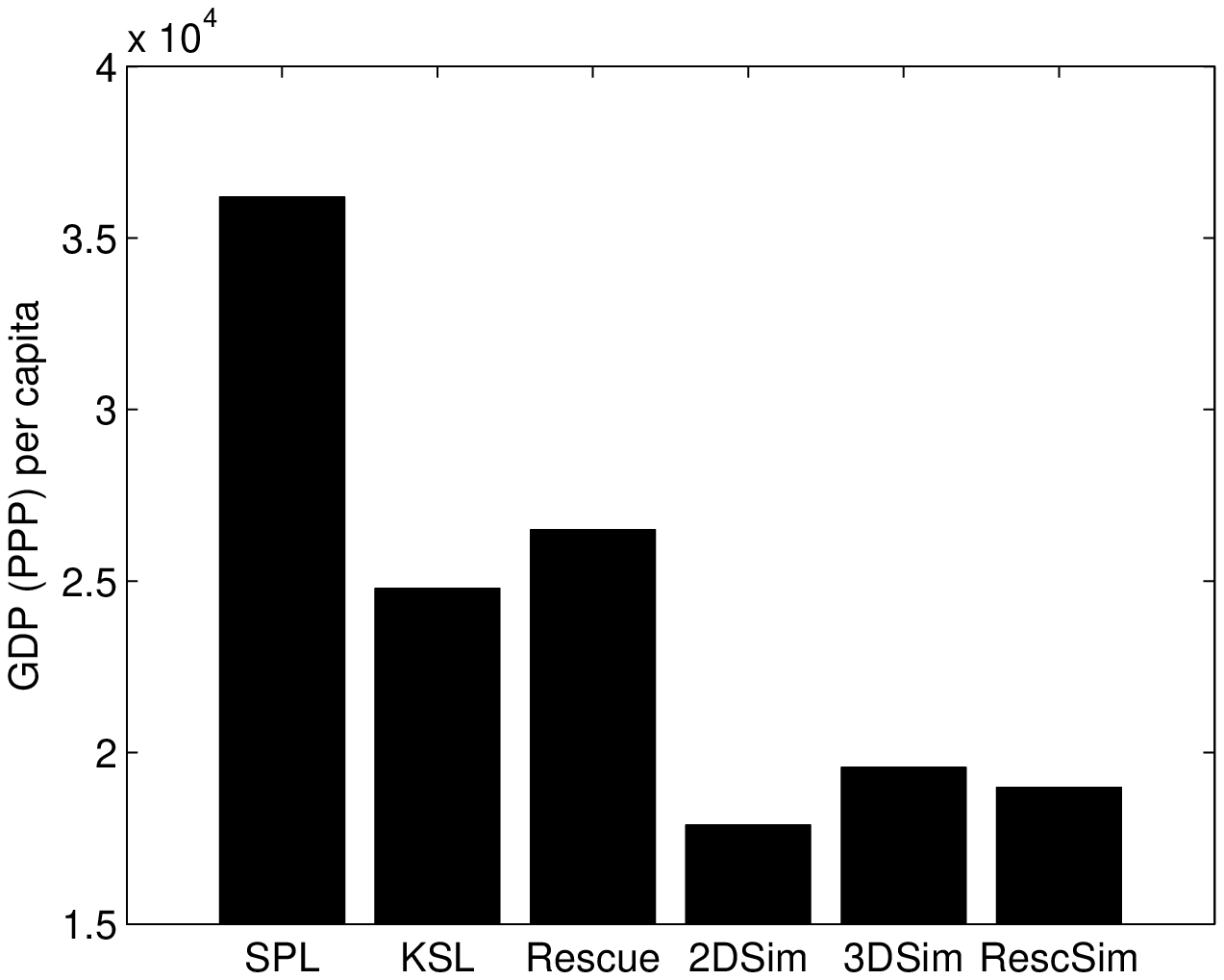}
\end{tabular}
\end{center}
\caption{PEoE (public expenditure on education as a percentage of GDP~\cite{world2012world}) and GDP/cap (gross domestic product at purchasing power parity per capita~\cite{world2012gdp}) for the home country of each participating RoboCup 2013 team, averaged over each of the six largest RoboCup leagues. Each of the three major simulation leagues (2D, 3D and rescue) exhibit significantly lower values than those requiring the purchase or development of physical robots. \label{fig:gdponed}}
\end{figure}

\mbox{}\newline\newline 
The physical robots required by non-simulation leagues remain particularly expensive. 
By removing these costs and those associated with robot repairs and transportation, the simulation leagues allow institutes with access to less funding to actively contribute to and participate in the RoboCup initiative. To quantify this claim, Fig.~\ref{fig:gdponed} presents the PEoE (public expenditure on education as a percentage of GDP~\cite{world2012world}) and GDP/cap (gross domestic product at purchasing power parity per capita~\cite{world2012gdp}) for the home country of each participating 
RoboCup 2013 team, averaged over each of the six largest RoboCup leagues. The countries participating in the standard platform league, which requires teams to field five Aldebaran Nao humanoid robots, have the highest average PEoE and GDP/cap of any league considered. The kid-sized humanoid and rescue leagues, each of which require the purchase or construction of physical robots, also involve countries with a high average PEoE and GDP/cap. Each of the three major simulation leagues (2D, 3D and rescue) exhibit significantly lower values, suggesting that the inclusion of simulation leagues supports financial inclusiveness within the competition.

\subsubsection{Statistically significant analyses by automated competition parallelism}
\mbox{}\newline\newline 
The automation of multiple parallel games makes RoboCupSoccer simulation leagues ideal platforms for analysing the complexities of complex team behaviours. Most team games and sports (both real and virtual) are characterised by rich, dynamic interactions that influence the contest outcome. As described by Vilar \emph{et al.}, ``quantitative analysis is increasingly being used
in team sports to better understand performance in these stylized, delineated, complex social systems"~\cite{vilar2013science}. Early examples of such quantitative analysis include \emph{sabermetrics}, which attempts to ``search for objective knowledge about baseball" by considering statistics of in-game activity~\cite{grabiner2004sabermetrics}. A recent study by Fewell \emph{et al.} involved the analysis of basketball games as networks, with properties including degree centrality, clustering, entropy and flow centrality calculating from measurements of ball position throughout the game~\cite{fewell2012basketball}. This idea was extended by Vilar \emph{et al.}, who considered the local dynamics of collective team behaviour to quantify how teams occupy sub-areas of the field as a function of ball position~\cite{vilar2013science}.
Recently, Cliff \emph{et al.} presented several information-theoretic methods of quantifying dynamic interactions in football games, using the RoboCupSoccer 2D simulation league as an experimental platform~\cite{cliff2013towards}.  

The ability to automate thousands of simulation league games allows for the analysis of competition formats to determine which best approximate the true performance rankings of competing teams. The selection of an appropriate competition format is critical for both the success and credibility of any competition. Unfortunately, this choice is not straightforward: The format must minimise randomness relative to the true performance ranking of teams while keeping the number of games to a minimum, both to satisfy time constraints and retain the interest of participants and spectators alike. Furthermore, maintaining competition interest introduces a number of constraints to competition formats: As an example, multiple games between the same two opponents (the obvious method of achieving a statistically significant ranking) should be avoided.

 The remainder of this paper quantifies the appropriateness of different tournament formats (a major consideration in many human sports) by determining the statistically significant performance rankings of 2012 and 2013 RoboCupSoccer 2D simulation teams. A new competition format is then proposed and verified by leveraging the automated parallelism facilitated by the 2D simulation league platform. In addition to demonstrating the utility of simulation leagues for statistical analysis of team sport outcomes given some system perturbation, it is anticipated that the adoption of the proposed format would improve the success and credibility of the RoboCupSoccer simulation leagues in future years.

\section{Previous competition formats}
\label{sec:previousstructures}

The following two competition formats were adopted by the RoboCupSoccer 2D simulation league in 2012 and 2013:

\begin{itemize}
\item In 2012, a total of 20 games were played to determine the final rank of the top 8 teams. Specifically, the top 4 teams played 6 games each (3 quarterfinal round-robin, 2 semifinal and 1 final/third place playoff), and the bottom 4 teams player 4 games each.
\item In 2013, a \emph{double-elimination} system was adopted, where a team ceases to be eligible to place first upon having lost 2 games~\cite{david1963method,edwards1996double}. A total of 16 games were played to determine the final rank of the top 8 teams. Specifically, 14 games were played in the double-elimination format (i.e. $2n-2, n=8$) in addition to 2 classification games.
\end{itemize}

Previously, it has been unclear whether this change in competition format improves the fairness and reproducibility of the final team rankings. In general, lack of reproducibility is due to non-transitivity of team performance (a well-known phenomena that occurs frequently in actual human team sports). This may be addressed by a round-robin competition (where all 28 possible pairs of teams play against one another), yet it is also unclear whether this increase in the number of games is guaranteed to improve ranking stability.

\section{Methods of ranking team performance}
\label{sec:methodsofranking}

Before evaluating different competition formats, it is necessary to establish a fair (i.e. statistically significant) ranking of the top 8 RoboCupSoccer 2D simulation league teams for 2012 and 2013. This was accomplished by conducting an 8-team round-robin for both years, where all 28 pairs of teams play approximately 1000 games against one another. In addition, two different schemes were considered for point calculation:

\begin{itemize}
\item \textbf{Continuous scheme:} Teams are ranked by sum of average points obtained against each opponent across all 1000 games.
\item \textbf{Discrete scheme:} Firstly, the average score between each pair of teams (across all 1000 games) is rounded to the nearest integer (e.g. ``1.9 : 1.2" is rounded to ``2 : 1"). Next, points are allocated for each pairing based on these rounded results: 3 for a win, 1 for a draw and 0 for a loss. Teams are then ranked by sum of these points received against each opponent.
\end{itemize}

The final rankings generated for 2012 and 2013 RoboCupSoccer 2D simulation league teams under these two schemes are presented in Sec.~\ref{ss:statsig}. Finally, in order to formally capture the overall difference between two rankings $\mathbf{r^a}$ and $\mathbf{r^b}$, the $L_1$ distance is utilised:

\begin{equation}
\label{eq:1}
    d_1(\mathbf{r^a}, \mathbf{r^b}) = \|\mathbf{r^a} - \mathbf{r^b}\|_1 = \sum_{i=1}^n |r^a_i-r^b_i| \ ,
\end{equation}

\noindent{where $i$ is the index of the $i$-th team in each ranking, $1 \leq i \leq 8$. The difference between rankings for different competition formats are presented in Sec.~\ref{ss:compstructs}.}

\section{Proposed competition format}
\label{sec:proposed}

Sec.~\ref{sec:methodsofranking} describes two schemes under which statistically significant rankings of RoboCupSoccer 2D simulation league teams can be achieved. However, it remains unclear whether the previously adopted competition formats are able to replicate these rankings with minimal noise for considerably fewer games, or whether a new format may achieve improved results in this regard. One possible format involves the following two steps:

\begin{itemize}
\item Firstly, a preliminary round-robin is conducted where 1 game is played for all 28 pairs of teams.
\item Following the rankings obtained in the previous step, 4 classification games are played: The final between the top 2 teams and playoffs between third and fourth, fifth and sixth, and seventh and eighth places. It is possibly to use the best-of-three format for each of these classification games.
\end{itemize}

\noindent{The 32 games required involved in this competition format could still fit readily in a 1-2 day time frame, particularly with 2 games running simultaneously as per RoboCup 2013.}

\section{Results}
\label{sec:results}

\subsection{Statistically significant rankings versus previous competition formats}
\label{ss:statsig}

Following iterated round-robin and two point calculation schemes described in Sec.~\ref{sec:methodsofranking}, statistically significant rankings were generated for the top 8 RoboCupSoccer 2D simulation league teams for 2012 and 2013. These results are presented below.

\subsubsection{RoboCup 2012 results}
\mbox{}\newline\newline
The final round-robin results of the top 8 teams for RoboCup 2012 are presented in Table~\ref{t1} and Table~\ref{t2}, for the continuous and discrete scoring schemes described in Sec.~\ref{sec:methodsofranking} respectively. Results are ordered according to actual performance at RoboCup 2012, $\mathbf{r^a}$.

Table~\ref{t2} presents the continuous (non-rounded) scores averaged across the approximately 1000 games for each pair in the round-robin, in addition to the points allocated according to the discretisation scheme (3 for a win, 1 for a draw and 0 for a loss). The tie-breaker is the rounded goal difference (not shown), which was used only to separate first place (WrightEagle, +39 points) from second (Helios, +26 points). The final ranking corresponds exactly with that generated under the continuous scheme, as presented in Table~\ref{t1}.

Despite the agreement between continuous and discrete scoring schemes, it is obvious that this ranking (generated from the results of approximately 28,000 games) disagrees significantly from the actual RoboCup 2012 results. This can be quantified using the distance metric defined in (\ref{eq:1}):

\begin{equation*}
    d_1(\mathbf{r^{a}}, \mathbf{r^{c}})_{2012} = |1 - 2| + |2 - 1| + |3 - 5| + |4 - 4| + |5 - 6| + |6 - 8| + |7 - 3| + |8 - 7| =  12,
\end{equation*}

\noindent{where $\mathbf{r^{a}}$ represents the actual RoboCup 2012 rankings and $\mathbf{r^{c}}$ represents the ranking generated under continuous scoring scheme round-robin. This large difference suggests that the 2012 competition format did not succeed in capturing the true team performance ranking.}

\begin{table}[t]
\begin{center}
\begin{tabular}{|c|c|c|c|c|c|c|c|c|c|c|c|c|}
 \hline
  \scriptsize{$\mathbf{r^a}$} & \scriptsize{Team} & \scriptsize{Helios} &  \scriptsize{Wright} & \scriptsize{Marlik} & \scriptsize{Gliders} & \scriptsize{GDUT} & \ \scriptsize{AUT} \ & \scriptsize{Yushan} &
 \scriptsize{RobOTTO} & \scriptsize{Points} & \scriptsize{Goal Diff} & \scriptsize{Rank, $\mathbf{r^c}$} \\ \hline
\scriptsize{1} & \scriptsize{Helios} & \cellcolor[gray]{0.4} & \scriptsize{1.397} & \scriptsize{2.442}  & \scriptsize{2.517}  & \scriptsize{2.948} & \scriptsize{2.970}  &  \scriptsize{2.880}  & \scriptsize{2.998}  & \scriptsize{18.152} & \scriptsize{+ 26.0} & \scriptsize{2} \\ \hline
\scriptsize{2} & \scriptsize{Wright} & \scriptsize{1.406}  & \cellcolor[gray]{0.4} & \scriptsize{2.792}  & \scriptsize{2.835}  & \scriptsize{2.900}  & \scriptsize{2.998}  & \scriptsize{2.970}  &  \scriptsize{2.998}  & \scriptsize{18.899} & \scriptsize{+ 38.7} &  \scriptsize{1}\\ \hline
\scriptsize{3} & \scriptsize{Marlik} & \scriptsize{0.309} & \scriptsize{0.129} & \cellcolor[gray]{0.4} & \scriptsize{1.147}  & \scriptsize{2.121}  & \scriptsize{2.804}  & \scriptsize{0.874}  & \scriptsize{2.615}  & \scriptsize{9.999} & \scriptsize{+ 0.3} & \scriptsize{5} \\ \hline
\scriptsize{4} & \scriptsize{Gliders} & \scriptsize{0.261} & \scriptsize{0.102} & \scriptsize{1.396} & \cellcolor[gray]{0.4} & \scriptsize{1.809}  & 	\scriptsize{2.957}  & \scriptsize{0.903}  & \scriptsize{2.863}  & \scriptsize{10.291} & \scriptsize{+ 3.4} & \scriptsize{4} \\ \hline
\scriptsize{5} & \scriptsize{GDUT}  & \scriptsize{0.029} & \scriptsize{0.074} & \scriptsize{0.633} & \scriptsize{0.960} & \cellcolor[gray]{0.4} & \scriptsize{2.955}  & 	\scriptsize{0.552}  & 	\scriptsize{2.597}  &  \scriptsize{7.800} & \scriptsize{- 6.0}  &  \scriptsize{6} \\ \hline
\scriptsize{6} & \scriptsize{AUT} & \scriptsize{0.007} & \scriptsize{0.001} & \scriptsize{0.107} & \scriptsize{0.026} & \scriptsize{0.024} & \cellcolor[gray]{0.4} & \scriptsize{0.003} & \scriptsize{0.209} &  \scriptsize{0.377} & \scriptsize{- 39.3} & \scriptsize{8} \\ \hline
\scriptsize{7} & \scriptsize{Yushan} & \scriptsize{0.084} & \scriptsize{0.021} & \scriptsize{1.822} & \scriptsize{1.875}  & \scriptsize{2.316}  & \scriptsize{2.994}  & \cellcolor[gray]{0.4} & \scriptsize{2.993}  & \scriptsize{12.105} & \scriptsize{+ 6.5} & \scriptsize{3} \\ \hline
\scriptsize{8} & \scriptsize{RobOTTO} & \scriptsize{0.001} &  \scriptsize{0.001} & \scriptsize{0.233} & \scriptsize{0.087} & \scriptsize{0.228} & \scriptsize{2.418}  & \scriptsize{0.005} & \cellcolor[gray]{0.4} & \scriptsize{2.973} & \scriptsize{- 29.6} & \scriptsize{7} \\ \hline
\end{tabular}
\end{center}
\caption{Round-robin results (average goals scored) for the top 8 teams from RoboCup 2012, ordered according to their final competition rank, $\mathbf{r^a}$. The final points for each team were determined by summing the average points scored against each opponent over approximately 1000 games, resulting in the round-robin with continuous point allocation scheme ranking, $\mathbf{r^c}$.}
\label{t1}

\begin{center}
\begin{tabular}{|c|c|c|c|c|c|c|c|c|c|c|c|c|}
 \hline
 & \scriptsize{Helios} &  \scriptsize{Wright} & \scriptsize{Marlik} & \scriptsize{Gliders} & \scriptsize{GDUT} & \ \scriptsize{AUT} \ &  \scriptsize{Yushan}  & \scriptsize{RobOTTO}  & \scriptsize{Goals} & \scriptsize{Points} & \scriptsize{$\mathbf{r^d}$} \\ \hline
\scriptsize{Helios} & \cellcolor[gray]{0.4} & \scriptsize{2.3 : 2.3} & \scriptsize{1.4 : 0.1}  & \scriptsize{1.6 : 0.1}  & \scriptsize{4.4 : 0.2} & \scriptsize{7.7 : 0.0}  &  \scriptsize{4.5 : 0.7}  & \scriptsize{7.6 : 0.1}  & \scriptsize{29.5 : 3.5} & \scriptsize{19} & \scriptsize{2} \\ \hline
\scriptsize{Wright} & \scriptsize{2.3 : 2.3}  & \cellcolor[gray]{0.4} & \scriptsize{3.2 : 0.3}  & \scriptsize{3.3 : 0.2}  & \scriptsize{5.8 :	1.2}  & \scriptsize{12.1 : 0.1}  & \scriptsize{7.2 : 1.0}  & \scriptsize{10.1 : 0.2}   & \scriptsize{44.0 : 5.3} & \scriptsize{19} & \scriptsize{1}\\ \hline
\scriptsize{Marlik} & \scriptsize{0.1 : 1.4} & \scriptsize{0.3 : 3.2} & \cellcolor[gray]{0.4} &  \scriptsize{0.46 : 0.56}  &  \scriptsize{1.4 : 0.4}  & \scriptsize{2.3 : 0.1}  & \scriptsize{0.7 : 1.2}  & \scriptsize{2.1 : 0.2}   & \scriptsize{7.4 : 7.1}  & \scriptsize{10} & \scriptsize{5}\\ \hline
\scriptsize{Gliders} & \scriptsize{0.1 : 1.6} & \scriptsize{0.2 : 3.3} &  \scriptsize{0.56 : 0.46}  & \cellcolor[gray]{0.4} & \scriptsize{1.9 : 1.2}  & 	\scriptsize{4.3 : 0.1}  & \scriptsize{1.4 : 2.2}  & \scriptsize{4.6 : 0.8}  & \scriptsize{13.1 : 9.7} & \scriptsize{12} & \scriptsize{4} \\ \hline
\scriptsize{GDUT}  & \scriptsize{0.2 : 4.4} & \scriptsize{1.2 : 5.8} & \scriptsize{0.4 : 1.4}  & \scriptsize{1.2 : 1.9} & \cellcolor[gray]{0.4} & \scriptsize{4.0 : 0.2}  & 	\scriptsize{2.0 : 3.9}  & 	\scriptsize{3.4 : 0.8}  &  \scriptsize{12.4 : 18.4}  & \scriptsize{6} & \scriptsize{6} \\ \hline
\scriptsize{AUT} & \scriptsize{0.0 : 7.7} & \scriptsize{0.1 : 12.1} & \scriptsize{0.1 : 2.3} & \scriptsize{0.1 : 4.3} & \scriptsize{0.2 : 4.0} & \cellcolor[gray]{0.4} & \scriptsize{0.1 : 7.1} & \scriptsize{0.7 : 3.1} & \scriptsize{1.3 : 40.6}  & \scriptsize{0} & \scriptsize{8}\\ \hline
\scriptsize{Yushan} & \scriptsize{0.7 : 4.5} & \scriptsize{1.0 : 7.2} & \scriptsize{1.2 : 0.7} & \scriptsize{2.2 : 1.4}  & \scriptsize{3.9 : 2.0}  & \scriptsize{7.1 : 0.1}  & \cellcolor[gray]{0.4} & \scriptsize{6.7 : 0.4}   &   \scriptsize{22.8 : 16.3}  & \scriptsize{13} & \scriptsize{3} \\ \hline
 \scriptsize{RobOTTO}  & \scriptsize{0.1 : 7.6} &  \scriptsize{0.2 : 10.1} & \scriptsize{0.2 : 2.1} & \scriptsize{0.8 : 4.6} & \scriptsize{0.8 : 3.4} & \scriptsize{3.1 : 0.7}  & \scriptsize{0.4 : 6.7} & \cellcolor[gray]{0.4}  & \scriptsize{5.6 : 35.2} & \scriptsize{3} & \scriptsize{7} \\ \hline
\end{tabular}
\end{center}
\caption{Round-robin results (average goals scored and discretised points allocated) for the top 8 teams from RoboCup 2012, ordered according to their final competition rank, $\mathbf{r^a}$. Discretised points are determined by calculating the average number of goals scored over approximately 1000 games rounded to the nearest integer, then awarding 3 points for a win, 1 point for a draw and 0 points for a loss. The resultant round-robin with discrete point allocation scheme ranking, $\mathbf{r^d}$, is equivalent to that generated under the continuous scheme.}
\label{t2}
\end{table}

\subsubsection{RoboCup 2013 results}

\begin{table}[t]
\begin{center}
\begin{tabular}{|c|c|c|c|c|c|c|c|c|c|c|c|c|}
 \hline
  \scriptsize{$\mathbf{r^a}$} & \scriptsize{Team} &  \scriptsize{Wright} &  \scriptsize{Helios} & \scriptsize{Yushan} & \scriptsize{Axiom} & \scriptsize{Gliders} & \ \scriptsize{Oxsy} \ &  \scriptsize{AUT}  &  \scriptsize{Cyrus}  & \scriptsize{Points} & \scriptsize{Goal Diff} & \scriptsize{Rank, $\mathbf{r^c}$} \\ \hline
\scriptsize{1} & \scriptsize{Wright} & \cellcolor[gray]{0.4} & \scriptsize{1.877} & \scriptsize{2.470}  & \scriptsize{2.880}  & \scriptsize{2.397} & \scriptsize{2.901}  &  \scriptsize{2.991}  & \scriptsize{2.792}  & \scriptsize{18.308} & \scriptsize{+ 22.5} & \scriptsize{1} \\ \hline
\scriptsize{2} & \scriptsize{Helios} & \scriptsize{0.883}  & \cellcolor[gray]{0.4} & \scriptsize{2.841}  & \scriptsize{2.940}  & \scriptsize{2.194}  & \scriptsize{2.343}  & \scriptsize{2.969}  & \scriptsize{2.767}  & \scriptsize{16.937} & \scriptsize{+ 14.9} &  \scriptsize{2}\\ \hline
\scriptsize{3} & \scriptsize{Yushan} & \scriptsize{0.406} & \scriptsize{0.093} & \cellcolor[gray]{0.4} & \scriptsize{2.506}  & \scriptsize{1.892}  & \scriptsize{1.557}  & \scriptsize{2.059}  & \scriptsize{0.921}  & \scriptsize{9.434} & \scriptsize{- 1.3} & \scriptsize{4} \\ \hline
\scriptsize{4} & \scriptsize{Axiom} & \scriptsize{0.072} & \scriptsize{0.042} & \scriptsize{0.367} & \cellcolor[gray]{0.4} & \scriptsize{0.590}  & 	\scriptsize{0.395}  & \scriptsize{1.224}  & \scriptsize{1.023}  & \scriptsize{3.713} & \scriptsize{- 14.5} & \scriptsize{8} \\ \hline
\scriptsize{5} & \scriptsize{Gliders}  & \scriptsize{0.437} & \scriptsize{0.490} & \scriptsize{0.884} & \scriptsize{2.249} & \cellcolor[gray]{0.4} & \scriptsize{1.612}  & 	\scriptsize{1.871}  & 	\scriptsize{0.828}  &  \scriptsize{8.371} & \scriptsize{- 2.0}  &  \scriptsize{6} \\ \hline
\scriptsize{6} & \scriptsize{Oxsy} & \scriptsize{0.065} & \scriptsize{0.385} & \scriptsize{1.159} & \scriptsize{2.437} & \scriptsize{1.105} & \cellcolor[gray]{0.4} & \scriptsize{2.225} & \scriptsize{2.167} &  \scriptsize{9.543} & \scriptsize{- 2.2} & \scriptsize{3} \\ \hline
\scriptsize{7} & \scriptsize{AUT} & \scriptsize{0.006} & \scriptsize{0.017} & \scriptsize{0.718} & \scriptsize{1.491}  & \scriptsize{0.878}  & \scriptsize{0.575}  & \cellcolor[gray]{0.4} & \scriptsize{0.731}  & \scriptsize{4.416} & \scriptsize{- 14.0} & \scriptsize{7} \\ \hline
\scriptsize{8} & \scriptsize{Cyrus} & \scriptsize{0.137} &  \scriptsize{0.136} & \scriptsize{1.791} & \scriptsize{1.740} & \scriptsize{1.926} & \scriptsize{0.632}  & \scriptsize{2.046} & \cellcolor[gray]{0.4} & \scriptsize{8.408} & \scriptsize{- 3.4} & \scriptsize{5} \\ \hline
\end{tabular}
\end{center}
\caption{Round-robin results (average goals scored) for the top 8 teams from RoboCup 2013, ordered according to their final competition rank, $\mathbf{r^a}$. The final points for each team were determined by summing the average points scored against each opponent over approximately 1000 games, resulting in the round-robin with continuous point allocation scheme ranking, $\mathbf{r^c}$.}

\label{t3}

\begin{center}
\begin{tabular}{|c|c|c|c|c|c|c|c|c|c|c|c|}
 \hline
 & \scriptsize{Wright} &  \scriptsize{Helios} & \scriptsize{Yushan} & \scriptsize{Axiom} & \scriptsize{Gliders} & \ \scriptsize{Oxsy} \ &  \scriptsize{AUT}  &  \scriptsize{Cyrus}  & \scriptsize{Goals} & \scriptsize{Points} & \scriptsize{$\mathbf{r^d}$} \\ \hline
\scriptsize{Wright} & \cellcolor[gray]{0.4} & \scriptsize{1.9 : 1.2} & \scriptsize{2.8 : 0.9}  & \scriptsize{4.9 : 0.3}  & \scriptsize{2.5 : 0.7} & \scriptsize{5.4 : 0.8}  &  \scriptsize{6.4 : 0.3}  & \scriptsize{3.4 : 0.6}  & \scriptsize{27.3 : 4.8} & \scriptsize{21} & \scriptsize{1} \\ \hline
\scriptsize{Helios} & \scriptsize{1.2 : 1.9}  & \cellcolor[gray]{0.4} & \scriptsize{2.8 : 0.2}  & \scriptsize{4.1 : 0.2}  & \scriptsize{1.2 : 0.2}  & \scriptsize{2.2 : 0.4}  & \scriptsize{4.1 : 0.1}  & \scriptsize{2.5 : 0.2}  & \scriptsize{18.1 : 3.2} & \scriptsize{18} &  \scriptsize{2}\\ \hline
\scriptsize{Yushan} & \scriptsize{0.9 : 2.8} & \scriptsize{0.2 : 2.8} & \cellcolor[gray]{0.4} & \scriptsize{2.7 : 0.8}  & \scriptsize{1.8 : 1.1}  & \scriptsize{1.4 : 1.2}  & \scriptsize{1.7 : 0.8}  & \scriptsize{0.9 : 1.4}  & \scriptsize{9.6 : 10.9} & \scriptsize{11} & \scriptsize{3} \\ \hline
\scriptsize{Axiom} & \scriptsize{0.3 : 4.9} & \scriptsize{0.2 : 4.1} & \scriptsize{0.8 : 2.7} & \cellcolor[gray]{0.4} & \scriptsize{1.0 : 2.3}  & 	\scriptsize{0.7 : 2.7}  & \scriptsize{0.9 : 1.1}  & \scriptsize{1.4 : 2.0}  & \scriptsize{5.3 : 19.8} & \scriptsize{1} & \scriptsize{8} \\ \hline
\scriptsize{Gliders}  & \scriptsize{0.7 : 2.5} & \scriptsize{0.2 : 1.2} & \scriptsize{1.1 : 1.8} & \scriptsize{2.3 : 1.0} & \cellcolor[gray]{0.4} & \scriptsize{1.3 : 1.0}  & 	\scriptsize{1.8 : 1.1}  & 	\scriptsize{0.9 : 1.7}  &  \scriptsize{8.3 : 10.3} & \scriptsize{7}  &  \scriptsize{6} \\ \hline
\scriptsize{Oxsy} & \scriptsize{0.8 : 5.4} & \scriptsize{0.4 : 2.2} & \scriptsize{1.2 : 1.4} & \scriptsize{2.7 : 0.7} & \scriptsize{1.0 : 1.3} & \cellcolor[gray]{0.4} & \scriptsize{2.3 : 0.8} &  \scriptsize{2.2 : 1.0}  &  \scriptsize{10.6 : 12.8} & \scriptsize{11} & \scriptsize{4} \\ \hline
\scriptsize{AUT} & \scriptsize{0.3 : 6.4} & \scriptsize{0.1 : 4.1} & \scriptsize{0.8 : 1.7} & \scriptsize{1.1 : 0.9}  & \scriptsize{1.1 : 1.8}  & \scriptsize{0.8 : 2.3}  & \cellcolor[gray]{0.4} & \scriptsize{0.8 : 1.8}  & \scriptsize{5.0 : 19.0} & \scriptsize{1} & \scriptsize{7} \\ \hline
\scriptsize{Cyrus} & \scriptsize{0.6 : 3.4} &  \scriptsize{0.2 : 2.5} & \scriptsize{1.4 : 0.9} & \scriptsize{2.0 : 1.4} & \scriptsize{1.7 : 0.9} &  \scriptsize{1.0 : 2.2}  & \scriptsize{1.8 : 0.8} & \cellcolor[gray]{0.4} & \scriptsize{8.7 : 12.1} & \scriptsize{10} & \scriptsize{5} \\ \hline
\end{tabular}
\end{center}
\caption{Round-robin results (average goals scored and discretised points allocated) for the top 8 teams from RoboCup 2012, ordered according to their final competition rank, $\mathbf{r^a}$. Discretised points are determined by calculating the average number of goals scored over approximately 1000 games rounded to the nearest integer, then awarding 3 points for a win, 1 point for a draw and 0 points for a loss. The tie-breaker is the total of rounded goal differences (not shown).  The resultant round-robin with discrete point allocation scheme ranking, $\mathbf{r^d}$, is slightly different to that generated under the continuous scheme.}
\label{t4}
\end{table}

\mbox{}\newline\newline 
The final round-robin results of the top 8 teams for RoboCup 2013 are presented in Table~\ref{t3} and Table~\ref{t4}, for the continuous and discrete scoring schemes described in Sec.~\ref{sec:methodsofranking} respectively. Results are ordered according to actual performance at RoboCup 2013, $\mathbf{r^a}$, and presented in the same format as Table~\ref{t1} and Table~\ref{t2} for RoboCup 2012.

Unlike RoboCup 2012, there is a slight disagreement between the rankings generated using continuous and discrete scoring schemes, with a swap between third and fourth teams. Again using the distance metric defined in (\ref{eq:1}), the difference between these rankings and the actual RoboCup 2013 results can be quantified:

\begin{equation*}
    d_1(\mathbf{r^{a}}, \mathbf{r^{c}})_{2013} = |1 - 1| + |2 - 2| + |3 - 4| + |4 - 8| + |5 - 6| + |6 - 3| + |7 - 7| + |8 - 5| =  12,
\end{equation*}
\begin{equation*}
    d_1(\mathbf{r^{a}}, \mathbf{r^{d}})_{2013} = |1 - 1| + |2 - 2| + |3 - 3| + |4 - 8| + |5 - 6| + |6 - 4| + |7 - 7| + |8 - 5| =  10,
\end{equation*}

\noindent{where $\mathbf{r^{a}}$ represents the actual RoboCup 2013 rankings, while $\mathbf{r^{c}}$ and $\mathbf{r^{d}}$ represent the ranking generated under continuous and discrete scoring schemes of round-robins respectively. It is evident that the 2013 double-elimination format yielded as much overall divergence as the 2012 single-elimination format, but with slightly fewer individual discrepancies. It is also clear that, given very small points differences between adjacent teams, it may be necessary to play classification games even after a statistically significant round-robin. It is therefore proposed that the format described in Sec.~\ref{sec:proposed} should improve reliability of the competition outcomes.

\subsection{Evaluation of proposed competition formats}
\label{ss:compstructs}

\begin{table}[t]
\begin{center}
\begin{tabular}{|c|c|c|c|c|c|c|c|c|c|c|c|c|c|}
 \hline
 & \scriptsize{Helios} &  \scriptsize{Wright} & \scriptsize{Marlik} & \scriptsize{Gliders} & \scriptsize{GDUT} & \ \scriptsize{AUT} \ &  \scriptsize{Yushan}  &  \scriptsize{RobOTTO}   & \scriptsize{Points} & \scriptsize{Rank} & \scriptsize{$\mathbf{r^{p}}$} \\ \hline
\scriptsize{Helios} & \cellcolor[gray]{0.4} & \scriptsize{4 : 1} & \scriptsize{4 : 0}  & \scriptsize{1 : 0}  & \scriptsize{4.4 : 0.2} & \scriptsize{1 : 0}  &  \scriptsize{4.5 : 0.7}  & \scriptsize{2 : 0}  &  \scriptsize{21} & \scriptsize{1} & \scriptsize{1} \\ \hline
\scriptsize{Wright} & \scriptsize{1 : 4}  & \cellcolor[gray]{0.4} & \scriptsize{2 : 1}  & \scriptsize{2 : 0}  & \scriptsize{5 :	1}  & \scriptsize{12.1 : 0.1}  & \scriptsize{6 : 1}  & \scriptsize{10.1 : 0.2}    & \scriptsize{18} & \scriptsize{2} & \scriptsize{2} \\ \hline
\scriptsize{Marlik} & \scriptsize{0 : 4} & \scriptsize{1 : 2} & \cellcolor[gray]{0.4} &  \scriptsize{1 : 0}  &  \scriptsize{1 : 0}  & \scriptsize{2.3 : 0.1}  & \scriptsize{1 : 1}  & \scriptsize{2.1 : 0.2}     & \scriptsize{13} & \scriptsize{3} & \scriptsize{4} \\ \hline
\scriptsize{Gliders} & \scriptsize{0 : 1} & \scriptsize{0 : 2} &  \scriptsize{0 : 1} & \cellcolor[gray]{0.4} & \scriptsize{1.9 : 1.2}  & 	\scriptsize{2 : 0}  & \scriptsize{1.4 : 2.2}  & \scriptsize{3 : 0}   & \scriptsize{9} & \scriptsize{5} & \scriptsize{5} \\ \hline
\scriptsize{GDUT}  & \scriptsize{0.2 : 4.4} & \scriptsize{1 : 5} & \scriptsize{0 : 1}  & \scriptsize{1.2 : 1.9} & \cellcolor[gray]{0.4} & \scriptsize{1 : 0}  & 	\scriptsize{3 : 2}  & 	\scriptsize{3.4 : 0.8}  &  \scriptsize{9} & \scriptsize{6} & \scriptsize{6} \\ \hline
\scriptsize{AUT} & \scriptsize{0 : 1} & \scriptsize{0.1 : 12.1} & \scriptsize{0.1 : 2.3} & \scriptsize{0 : 2} & \scriptsize{0 : 1} & \cellcolor[gray]{0.4} & \scriptsize{0.1 : 7.1} & \scriptsize{1 : 0} &  \scriptsize{3} & \scriptsize{7} & \scriptsize{8} \\ \hline
\scriptsize{Yushan} & \scriptsize{0.7 : 4.5} & \scriptsize{1 : 6} & \scriptsize{1 : 1} & \scriptsize{2.2 : 1.4}  & \scriptsize{2 : 3}  & \scriptsize{7.1 : 0.1}  & \cellcolor[gray]{0.4} & \scriptsize{3 : 1}   & \scriptsize{10} & \scriptsize{4} & \scriptsize{3} \\ \hline
\scriptsize{RobOTTO} & \scriptsize{0 : 2} &  \scriptsize{0.2 : 10.1} & \scriptsize{0.2 : 2.1} & \scriptsize{0 : 3} & \scriptsize{0.8 : 3.4} & \scriptsize{0 : 1}  & \scriptsize{1 : 3} & \cellcolor[gray]{0.4}  & \scriptsize{0} & \scriptsize{8} & \scriptsize{7} \\ \hline
\end{tabular}
\end{center}
\caption{Combined actual and average results for the top 8 teams from RoboCup 2012, ordered according to their final competition rank. Each goal difference represents the actual (integer) game results from RoboCup 2012 where possible. As this previous format does not necessarily require all pairs of teams to play against one another, some of these results are not available: In these cases, the average (continuous-valued) scores from Table~\ref{t2} were utilised. Using these results, it is possible to infer the final ranking, $\mathbf{r^p}$, for RoboCup 2012 under the competition format proposed in Sec.~\ref{sec:proposed}.}
\label{t5}
\end{table}

In order to evaluate the proposed competition format described in Sec.~\ref{sec:proposed}, the actual game results from RoboCup 2012 and 2013 were used where possible. As these previous formats do not necessarily require all pairs of teams to play against one another, some of these results are not available: In these cases, the average scores from Table~\ref{t2} and Table~\ref{t4} were utilised for RoboCup 2012 and 2013 respectively.

Using these results, it is possible to infer final rankings for RoboCup 2012 and 2013 under the proposed competition format. These results are presented below.

\subsubsection{RoboCup 2012 results}
\mbox{}\newline\newline
The combined actual and average results of top 8 teams from RoboCup 2012 are presented in Table~\ref{t5}, in addition to the inferred final ranking, $\mathbf{r^p}$, for RoboCup 2012 under the competition format proposed in Sec.~\ref{sec:proposed}. Using the distance metric defined in (\ref{eq:1}), the difference between $\mathbf{r^p}$ and the ranking generated from the 28,000 game round-robin, $\mathbf{r^c}$, can be quantified:

\begin{equation*}
    d_1(\mathbf{r^{p}}, \mathbf{r^{c}})_{2012} = |1 - 2| + |2 - 1| + |4 - 5| + |5 - 4| + |6 - 6| + |8 - 8| + |3 - 3| + |7 - 7| =  4.
\end{equation*}

\noindent{This is a considerably smaller difference than the 12 produced under the actual RoboCup 2012 format, suggesting that the proposed format better captures the true team performance ranking. Furthermore, this result is achieved using a majority of actual game results (i.e. 18 from 28 pairs, with only 10 using the averages from Table~\ref{t2}).


\subsubsection{RoboCup 2013 results}
\begin{table}[t]
\begin{center}
\begin{tabular}{|c|c|c|c|c|c|c|c|c|c|c|c|c|c|}
 \hline
 & \scriptsize{Wright} &  \scriptsize{Helios} & \scriptsize{Yushan} & \scriptsize{Axiom} & \scriptsize{Gliders} & \ \scriptsize{Oxsy} \ & \scriptsize{AUT} &  \scriptsize{Cyrus}   & \scriptsize{Points} & \scriptsize{Rank} & \scriptsize{$\mathbf{r^{p}}$}  \\ \hline
\scriptsize{Wright} & \cellcolor[gray]{0.4} & \scriptsize{3 : 1} & \scriptsize{2.8 : 0.9}  & \scriptsize{4.9 : 0.3}  & \scriptsize{2.5 : 0.7} & \scriptsize{5 : 3}  &  \scriptsize{6.4 : 0.3}  & \scriptsize{7 : 0}  &  \scriptsize{21} & \scriptsize{1} & \scriptsize{1} \\ \hline
\scriptsize{Helios} & \scriptsize{1 : 3}  & \cellcolor[gray]{0.4} & \scriptsize{2 : 0}  & \scriptsize{4.1 : 0.2}  & \scriptsize{2 : 0}  & \scriptsize{2.2 : 0.4}  & \scriptsize{4.1 : 0.1}  & \scriptsize{4 : 0}  &  \scriptsize{18} &  \scriptsize{2} & \scriptsize{2} \\ \hline
\scriptsize{Yushan} & \scriptsize{0.9 : 2.8} & \scriptsize{0 : 2} & \cellcolor[gray]{0.4} & \scriptsize{4 : 1}  & \scriptsize{0 : 1}  & \scriptsize{3 : 0}  & \scriptsize{1.7 : 0.8}  & \scriptsize{2 : 0}  &  \scriptsize{12} & \scriptsize{3} & \scriptsize{3} \\ \hline
\scriptsize{Axiom} & \scriptsize{0.3 : 4.9} & \scriptsize{0.2 : 4.1} & \scriptsize{1 : 4} & \cellcolor[gray]{0.4} & \scriptsize{3 : 3}  & 	\scriptsize{1 : 6}  & \scriptsize{2 : 1}  & \scriptsize{1.4 : 2.0}  &  \scriptsize{4} & \scriptsize{7} & \scriptsize{8} \\ \hline
\scriptsize{Gliders}  & \scriptsize{0.7 : 2.5} & \scriptsize{0 : 2} & \scriptsize{1 : 0} & \scriptsize{3 : 3} & \cellcolor[gray]{0.4} & \scriptsize{4 : 0}  & 	\scriptsize{1.8 : 1.1}  & 	\scriptsize{0.9 : 1.7}  &   \scriptsize{10}  &  \scriptsize{4} & \scriptsize{4} \\ \hline
\scriptsize{Oxsy} & \scriptsize{3 : 5} & \scriptsize{0.4 : 2.2} & \scriptsize{0 : 3} & \scriptsize{6 : 1} & \scriptsize{0 : 4} & \cellcolor[gray]{0.4} & \scriptsize{2.3 : 0.8} &  \scriptsize{2.2 : 1.0}  &   \scriptsize{9} & \scriptsize{5} & \scriptsize{5} \\ \hline
\scriptsize{AUT} & \scriptsize{0 : 7} & \scriptsize{0.1 : 4.1} & \scriptsize{0.8 : 1.7} & \scriptsize{1 : 2}  & \scriptsize{1.1 : 1.8}  & \scriptsize{0.8 : 2.3}  & \cellcolor[gray]{0.4} & \scriptsize{3 : 1}  &  \scriptsize{3} & \scriptsize{8} & \scriptsize{7} \\ \hline
\scriptsize{Cyrus} & \scriptsize{0.6 : 3.4} &  \scriptsize{0 : 4} & \scriptsize{0 : 2} & \scriptsize{2.0 : 1.4} & \scriptsize{1.7 : 0.9} &  \scriptsize{1.0 : 2.2} & \scriptsize{1 : 3} & \cellcolor[gray]{0.4} & \scriptsize{6} & \scriptsize{6} & \scriptsize{6} \\ \hline
\end{tabular}
\end{center}
\caption{Combined actual and average results for the top 8 teams from RoboCup 2013, ordered according to their final competition rank. Each goal difference represents the actual (integer) game results from RoboCup 2013 where possible. As this previous format does not necessarily require all pairs of teams to play against one another, some of these results are not available: In these cases, the average (continuous-valued) scores from Table~\ref{t4} were utilised. Using these results, it is possible to infer the final ranking, $\mathbf{r^p}$, for RoboCup 2013 under the competition format proposed in Sec.~\ref{sec:proposed}.}
\label{t6}
\end{table}
\mbox{}\newline\newline 
The combined actual and average results of top 8 teams from RoboCup 2013 are presented in Table~\ref{t6}, in addition to the inferred final ranking, $\mathbf{r^p}$, for RoboCup 2013 under the competition format proposed in Sec.~\ref{sec:proposed}. Again using the distance metric defined in (\ref{eq:1}), the difference between $\mathbf{r^p}$ and the ranking generated from the 28,000 game round-robin, $\mathbf{r^c}$ or $\mathbf{r^d}$, can be quantified:

\begin{equation*}
    d_1(\mathbf{r^{p}}, \mathbf{r^{c}})_{2013} = |1 - 1| + |2 - 2| + |3 - 4| + |8 - 8| + |4 - 6| + |5 - 3| + |7 - 7| + |6 - 5| =  6
\end{equation*}
\begin{equation*}
    d_1(\mathbf{r^{p}}, \mathbf{r^{d}})_{2013} = |1 - 1| + |2 - 2| + |3 - 3| + |8 - 8| + |4 - 6| + |5 - 4| + |7 - 7| + |6 - 5| =  4
\end{equation*}

\noindent{Similarly to the results for 2012, these are considerably smaller differences than the 12 (or 10) produced under the actual RoboCup 2013 format, providing further evidence that the proposed format better captures the true team performance ranking. Again, this result is achieved using a majority of actual game results (i.e. 15 from 28 pairs, with only 13 using the averages from Table~\ref{t4}).

\newpage

\subsubsection{Statistical validation}
\mbox{}\newline\newline
To statistically validate that the proposed competition format is significantly more appropriate than those adopted at RoboCup 2012 and 2013, 10,000 tournaments were generated for each format by randomly sampling game results from the 28,000 game round-robin. For each tournament, the $L_1$ distance, $d_1(\mathbf{r^a}, \mathbf{r^b})$ (see eq. (\ref{eq:1})), was calculated to capture the discrepancy between the tournament and true team rankings. These results are presented in Fig.~\ref{fig:comparison} for the top 8 teams from RoboCup 2012 (a) and 2013 (b). In both cases, it is evident that the proposed format yields more statistically robust rankings (i.e. smaller $L_1$ distance) than the formats adopted in RoboCup 2012 (second best) and 2013 (worst).

\begin{figure}
\begin{center}
\begin{tabular}{>{\centering}p{5.8cm}<{\centering} >{\centering}p{5.8cm}<{\centering}}
\includegraphics[width=5.8cm]{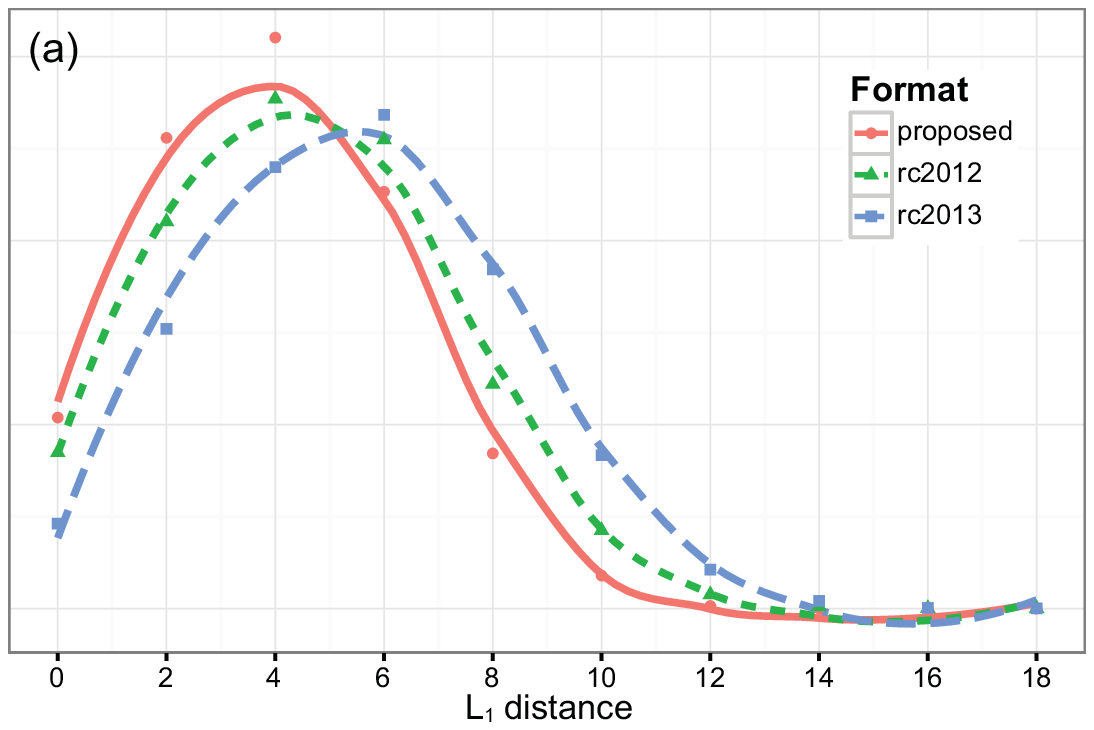} & \includegraphics[width=5.8cm]{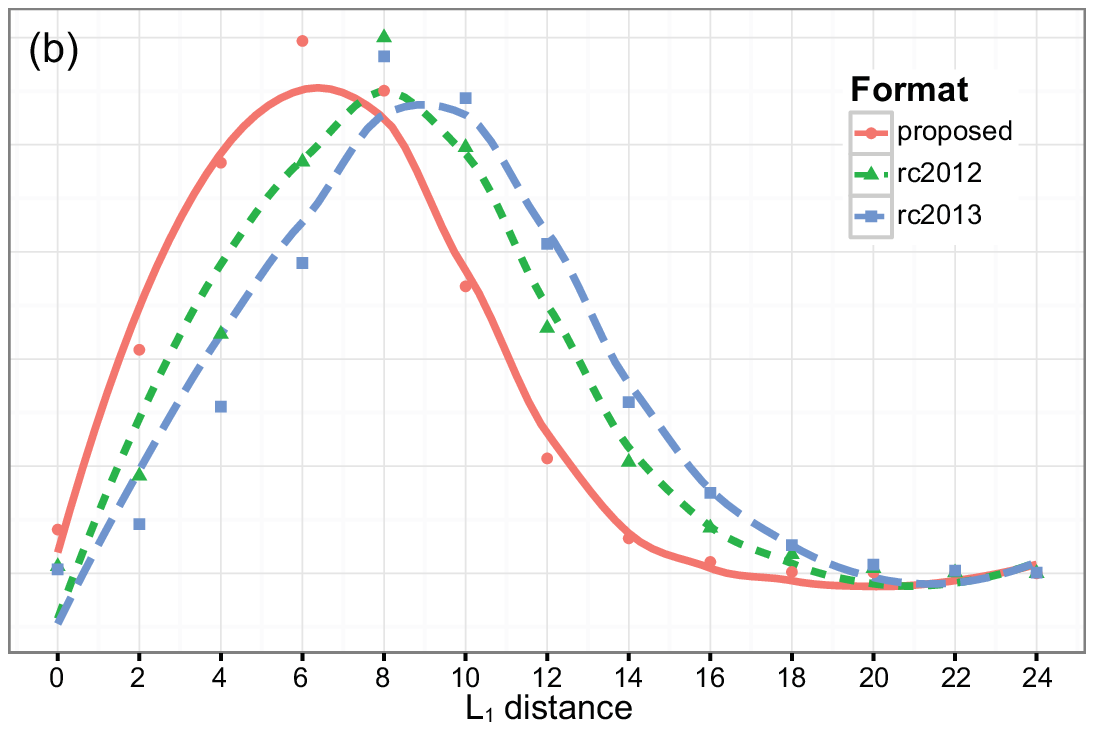}
\end{tabular}
\end{center}
\caption{Discrepancy between tournament and true team rankings, captured as an $L_1$ distance (see eq. (\ref{eq:1})), for 10,000 randomly-generated tournaments structured according to the three considered formats. It is evident that the proposed format (red) yields more statistically robust rankings (i.e. smaller $L_1$ distance) than the formats adopted in RoboCup 2012 (second best, green) and 2013 (worst, blue), considering the top 8 teams from both RoboCup 2012 (a) and 2013 (b). \label{fig:comparison}}
\end{figure}

\section{Conclusions}\label{conclusions}

The selection of an appropriate competition
format is critical for both the success and credibility of any competition. This is particularly true in the RoboCupSoccer 2D simulation league, which provides an ideal computational platform for examining different formats by facilitating automated parallel execution of a statistically significant number of games.

A 28,000 game round-robin competition was conducted between the top 8 2D simulation league teams from both RoboCup 2012 and 2013. The difference between the resultant rankings was calculated relative to the actual results of RoboCup 2012 and 2013 (12 and 12 respectively) and compared to those that would have resulted under a proposed new format (4 and 6 respectively). This suggests a significant reduction in randomness relative to true team performance rankings while only requiring the number of games to be increased to 32; a number that would still fit readily in a 1-2 day time frame, particularly utilising the round-robin parallelism enabled by the stable 2D simulation platform.

The RoboCup ``Millennium Challenge" requires robots to exhibit both physical and strategic prowess, necessitating the decomposition of the larger problem into both physical robot and simulation leagues. Although often overlooked, the simulation leagues contribute significantly to this goal, both through improving financial inclusiveness of competing nations and providing a stable platform for statistically significant analysis of team behaviour and competition format. In addition to highlighting the latter of these contributions, it is anticipated that the introduction of the proposed new format will improve the reliability of final competition rankings and consequently success and credibility of the RoboCupSoccer simulation leagues in future years.



\bibliographystyle{splncs}
\bibliography{refs}

\end{document}